\documentclass[10pt,final,journal]{IEEEtran}

\ifCLASSINFOpdf
  \usepackage{graphicx}
  \usepackage{epstopdf}
\else
\fi
\usepackage{amsmath}
\usepackage{amssymb}
\usepackage{pgfplots} 
\usepackage{pgfgantt}
\usepackage{pdflscape}
\pgfplotsset{compat=newest} 
\pgfplotsset{plot coordinates/math parser=false}

\usepackage{color}
\newcommand{\Fig}[1]{\mbox{Fig.~\ref{#1}}}
\newcommand{\eq}[1]{\mbox{Eq.~(\ref{#1})}}
\newcommand{\mb}[1]{\ensuremath\mathbf{#1}}

\begin{document}
%
\title{Tight Bounds for Uncertain Time-Correlated Errors with Gauss-Markov Structure}
%
%
%
%
\author{Omar~Garc\'ia~Crespillo,~\IEEEmembership{Member,~IEEE,}
        Mathieu~Joerger,~\IEEEmembership{Member,~IEEE,}
        and~Steve~Langel,~\IEEEmembership{Member,~IEEE}
\thanks{O.~Garc\'ia~Crespillo is with the Navigation Department of the German Aerospace Center (DLR), Oberpfaffenhofen, Germany. He is also a PhD student at Swiss Federal Institute of Technology in Lausanne (EPFL).\newline E-mail: Omar.GarciaCrespillo@dlr.de}
\thanks{M. Joerger is with the Department of Aerospace and Ocean Engineering, Virginia Tech.}
\thanks{S. Langel is with the Department of Communications, SIGINT and PNT, The MITRE Corportation.}
\thanks{}
}
\maketitle
\begin{abstract}
Safety-critical navigation applications require that estimation errors be  reliably quantified and bounded. This can be challenging for linear dynamic systems if the process noise or measurement errors have uncertain time correlation. 
In many systems (e.g., in satellite-based or inertial navigation systems), there are sources of time-correlated sensor errors that can be well modeled using Gauss-Markov processes (GMP). 
However, uncertainty in the GMP parameters, particularly in the correlation time constant, can cause misleading error estimation.
In this paper, we develop new time-correlated models that ensure tight upper bounds on the estimation error variance, assuming that the actual error is a stationary GMP with a time constant that is only known to reside within an interval. We first use frequency-domain analysis to derive a stationary GMP model both in continuous and discrete time domain, which outperforms models previously described in the literature. Then, we achieve an even tighter estimation error bound using a non-stationary GMP model, for which we determine the minimum initial variance that guarantees bounding conditions. In both cases, the model can easily be implemented in a linear estimator like a Kalman filter. 
\end{abstract}
%
\IEEEpeerreviewmaketitle
\section{Introduction}
\IEEEPARstart{S}{afety} and liability-critical applications require a guaranteed bound on the estimation error, even when process and measurement noise cannot be precisely characterized.
The concept of cumulative distribution function (CDF) overbounding  supports safe error quantification in the context of Global Navigation Satellite Systems (GNSS) positioning~\cite{Lee2009_SigmaOverbounding,Rife2012_OverboundRevisited}. However, CDF overbounding is designed to be used only for snapshot estimators such as least-squares estimators \cite{Decleene2000}, \cite{Rife2006_PairedOverbounding_TAES}, \cite{Blanch2019_2stepBound}. That is, it does not directly apply to linear dynamic systems because it does not account for measurement error correlation \emph{over time}.
New navigation applications are emerging that require the use of Kalman filters (KF) or other sequential or fixed-lag estimators to meet stringent requirements and to incorporate information from other sensors (e.g. from inertial navigation systems or INS).

Robust estimation approaches were developed to address  model uncertainty in linear dynamic systems. For instance, the optimization of a scaling parameter to bound the estimation error using a discrete-time KF is proposed in \cite{Xie1994_RobustKF}. This approach can present limitations due to the need to perform an optimization process at each time step, and it only considers uncertainty in the process and measurement design matrices, not in the noise terms that are of main interest in this paper. In \cite{Petersen1994_GuaranteedCost}, the authors also aim at  guaranteed cost filtering under system uncertainty, but noise structure uncertainty is not considered. 
Other robust filters use norm-bounded cost functions based on H$\infty$ or H2/H$\infty$~\cite{Bernstein1989_L1asHinf,Yang2005_Hinf} or use cost functions based on M-estimators~\cite{Gandhi2010_M-estimatorKF}. They have been implemented in navigation applications and show great potential~\cite{GarciaCrespillo2018}. But, they do  not allow for rigorous estimation error bounding, which makes makes them unfit for safety-critical applications.

The authors in \cite{Rife2007_SymmOverbounding, Pulford2008} provided bounds on the error of linear systems with spherically symmetric time correlated measurement errors. In \cite{Langel2014_AIAA} a bounding approach is proposed when the auto-correlation function of measurement or process noises can be upper and lower bounded. While these methods do not require any knowledge about the structure of the noises, they require evaluating the impact of all previous time epochs in a batch processing scheme. This is a limitation for real-time systems since the required operations and memory allocation grow fast as time passes.
In \cite{Langel2020b}, the authors showed that the true KF estimation error covariance could be upper bounded if the Power Spectral Density (PSD) of the measurement or process error model upper bounded that of the actual time-correlated sensor errors for any frequency.

Realistic time-correlated errors can have complex time correlation structures.  To model these structures in practical applications, Gauss-Markov Processes (GMP) are widely employed both because they can be reasonably accurate and because they have a simple two-parameter formulation. GMPs can easily be incorporated in a KF by state augmentation.

In \cite{Langel2019, Tupysev2009}, a stationary GMP model was derived by KF sensitivity analysis. A tighter non-stationary GMP model was also provided in \cite{Langel2019}.
In~\cite{GarciaCrespillo2020b_OverboundGNSSINS}, this GMP model was implemented in a GNSS/INS KF for an aircraft landing application. 
Also in \cite{GarciaCrespillo2020b_OverboundGNSSINS}, by graphical inspection of the GMP model's Allan variance, the authors made the conjecture that a GMP model  existed, which would provide a tighter bound on the estimation error variance.

In this paper, we derive the tightest stationary GMP models that upper bounds the estimation errors in the presence of measurement or process noises with GMP structure but uncertain variance and time constant. The GMP model is found in the PSD domain both in the continuous and discrete time domain. Then, we tighten this bound considering a non-stationary GMP whose initial variance we can reduce. The new bound is tighter during the transient phase of the non-stationary process, and remains bounding at steady state.

This paper is organized as follows. In Section~\ref{sec:stationaryBoundCont}, we derive the parameters of a new stationary GMP bound in the continuous-time domain and represent its PSD. In Section~\ref{sec:stationaryBoundDis}, we present the stationary GMP model derived in the discrete-time domain. In Section~\ref{sec:NonstationaryBound}, we derive the minimum initial variance of the non-stationary GMP that guarantees an upper bound on the estimation error variance. In Section~\ref{sec:example}, we evaluate the bounds for an example KF implementation. Concluding remarks are given in Section~\ref{sec:Conclusion}.
\section{Stationary Continuous-Time GMP Model}\label{sec:stationaryBoundCont}
\subsection{Problem Formulation}
We consider a linear dynamic system~(LDS) described by a continuous-time system model with a discrete-time measurement model:
\begin{align}
    \dot{\mb{x}}(t) &= \mb{F}(t)\mb{x}(t) + \mb{w}(t), \\
    \mb{z}_k &= \mb{H}_k\mb{x}_k + \mb{v}_k,
\end{align}
where $\mb{x}$ is the state vector, $\mb{F}$ is the state transition matrix, $\mb{z}$ is the observation or measurement vector, and $\mb{w}(t)$ and $\mb{v}_k$ are time-correlated process and measurement noise vectors. One solution to account for the time-correlated component of the noise vectors is to augment the LDS with states corresponding to the time-correlated components of $\mb{w}(t)$ and $\mb{v}_k$. 
In many practical applications, the augmented state is modeled as a GMP. For instance, in \cite{RTCA-DO-229D} it is specified to use a GMP to model the tropospheric and satellite ephemeris and clock errors in GPS/INS tight integration. In the KF, the variance of the GMP process noises represent their PSD.
 
In~\cite{Langel2020}, formal proof is given that using an error model whose PSD is upper bounding that of the empirical error would guarantee a bound on the estimation error variance for any linear estimator, be it batch or Kalman filter. 
A GMP $a$ can be expressed as:
\begin{align}
    \dot{a}(t) = \frac{1}{\tau}a(t) + \sqrt{\frac{2\sigma^2}{\tau}}w(t),\;\text{with}\;w(t) \sim \mathcal{N}(0,1)
\end{align}
where $\tau\in\mathbb{R}>0$ is the GMP correlation time constant and $\sigma^2\in\mathbb{R}\geq 0$ is the stationary process variance.
In the first part of this paper, we seek to find a GMP model whose PSD upper bounds that of an actual GMP with uncertain time constant $\tau \in [\tau_\text{min},\tau_\text{max}]$, for all frequencies. 
Since the spectrum of the real process $a$ is an even function, we only need to bound the PSD over $[0,\infty)$:
\begin{align}\label{eq:mainCondition}
    \hat{S}(\omega) \geq S(\omega), \forall \omega\in[0,\infty),
\end{align}
where $\hat{S}$ is the bounding GMP PSD, $S$ is the PSD of the actual GMP, and the angular frequency in [rad/s] is $\omega=2\pi f$ with $f$ being the linear frequency in [Hz].

Furthermore, we want to find the tightest possible bound $\hat{S}$ in order to minimize the total net power of the process, i.e., the variance of the GMP model, and its resulting contribution to the linear estimation process.

\subsection{Continuous-Time Model Derivation}
\begin{figure}[t]
    \newlength\fwidth
    \setlength\fwidth{0.85\linewidth}
    \centering
    \input{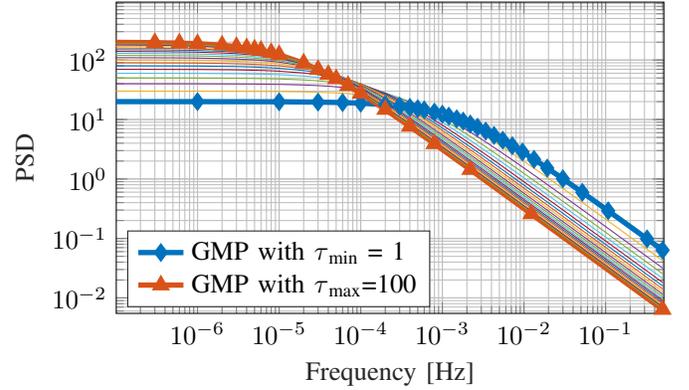}
    \caption{Power Spectral Density (PSD) of GM processes with with $\sigma^2 = 1$ and $\tau \in [10, 100] s$.}
    \label{fig:psdNobound}
\end{figure}

The spectral density of a GMP $a$ can be expressed as~\cite{Brown2012}:
\begin{align}\label{eq:GMP_PSD}
    S(\omega) = \frac{2\sigma^2/\tau}{\omega^2 + (1/\tau)^2},
\end{align}
where $\sigma^2$ is given  (or, an upper bound on $\sigma^2$ may be used), and we know that the time-constant of the GMP exists within a range $\tau \in [\tau_\text{min}, \tau_\text{max}]$. The time constant of the GMP shapes its PSD. For example, in \Fig{fig:psdNobound}, we show PSD curves for GMPs with time constants ranging from 10 to 100 seconds.

In \Fig{fig:psdNobound}, actual GMPs with different time constants cross at different frequencies. As explained in \cite{Langel2019}, this illustrates the fact that a KF designed assuming the maximum measurement and process error time correlation produces an estimation error variance that does not necessarily bound its actual value at any time (or at any frequency) is incorrect.

We are interested in finding a GMP model with correlation time constant $\hat{\tau} \in \mathbb{R}$ and inflation factor $k\in \mathbb{R}$ such that the GMP model variance $\hat{\sigma}^2 = k \sigma^2$ is obtained as a function of the underlying process variance.

According to \eq{eq:GMP_PSD} and \Fig{fig:psdNobound} the shape of the GMP PSD is driven by the relative values of $\omega^2$ and $(1/\tau)^2$. We can therefore analyze two important cases: Case 1, when $\omega<<1/\tau$, i.e., when $\omega\to 0$; and Case 2, when $\omega >> 1/\tau$ where $S(\omega)$ follows its behavior as $\omega$ approaches infinity. 

In Case 1, we can write:
\begin{align}
    S(\omega\to 0) =\lim_{\omega\to 0} \frac{2\sigma^2/\tau}{\omega^2 + (1/\tau)^2} =
    \frac{2\sigma^2/\tau}{(1/\tau)^2} = 2\sigma^2\tau. \label{eq:cond1}
\end{align}
It is worth noting that in \eq{eq:cond1}, the value of $\tau$ that produces the PSD with the highest value is $\tau_\text{max}$. So in order to satisfy \eq{eq:mainCondition}, our first condition is:
\begin{align}
    \label{eq:Cond1}
    \hat{S}(\omega\to 0) &\geq S(\omega\to 0), \nonumber \\
    2k\sigma^2\hat{\tau} &\geq 2\sigma^2\tau_\text{max}, \nonumber\\
    \text{Constraint 1:}\;\;\;\;\;\; k\hat{\tau} &\geq \tau_\text{max}.
\end{align}

In Case 2 where $\omega^2 >> (1/\tau)^2$, we have:
\begin{align}
    S(\omega>>1/\tau) \leq \frac{2\sigma^2/\tau}{\omega^2}.
\end{align}
In this expression, the highest PSD value is obtained when $\tau = \tau_\text{min}$. This leads to our second condition:
\begin{align}
   \label{eq:Cond2}
    \hat{S}(\omega >>1/\tau_\text{min}) &\geq S(\omega>>1/\tau_\text{min}), \nonumber \\
    \frac{2k\sigma^2/\hat{\tau}}{\omega^2} &\geq \frac{2\sigma^2/\tau_\text{min}}{\omega^2}, \nonumber \\
    \text{Constraint 2:}\;\;\;\;\;\; \frac{k}{\hat{\tau}} &\geq \frac{1}{\tau_\text{min}}.
\end{align}

\begin{figure}[t]
    \centering
    \includegraphics[width=\linewidth]{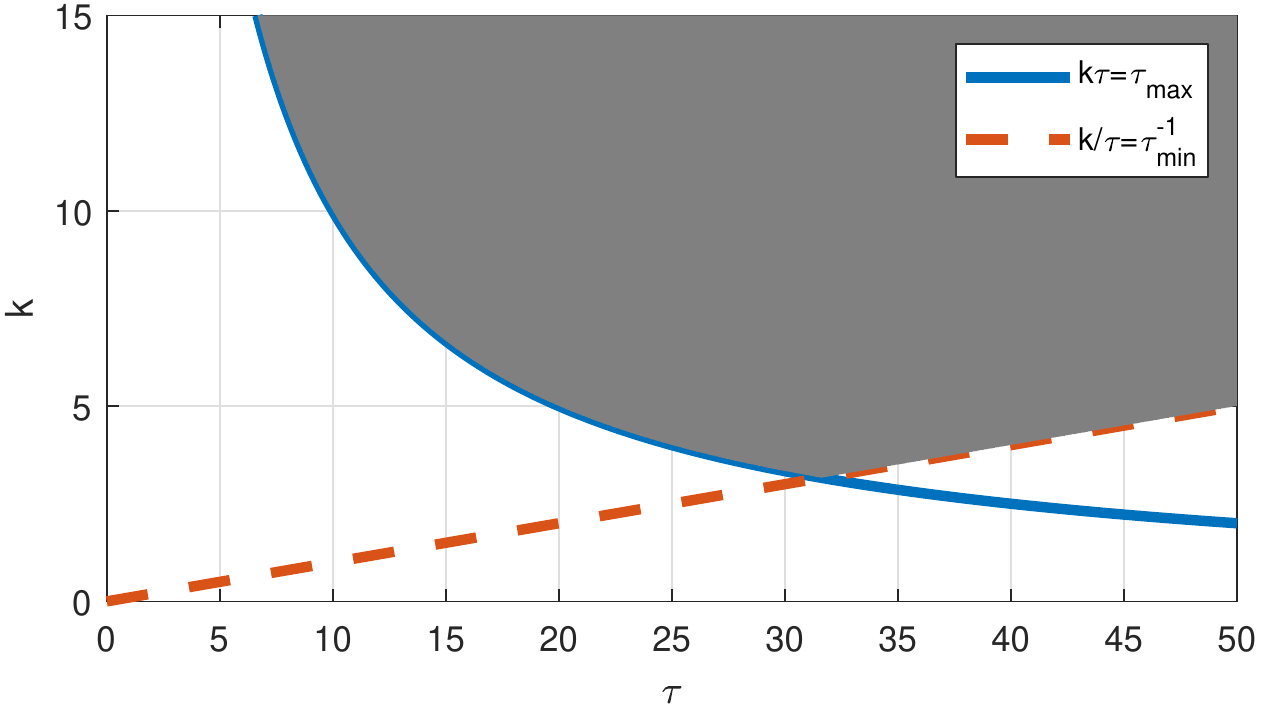}
    \caption{Optimization solution area (in gray) for $\sigma^2 = 1$ and $\tau \in [10,100] $s.}
    \label{fig:OptArea}
\end{figure}

The two conditions constrain the range of admissible values for $k$ and $\hat{\tau}$ ensuring that \eq{eq:mainCondition} is satisfied. The area of possible solutions is shown in \Fig{fig:OptArea}.
Out of all candidate solutions, we want to find the one that produces the tightest $\hat{S}$-bound. The tightest bound is the one that minimizes the total net power of the process, which is also its variance:
\begin{align}
\min \frac{1}{\pi}\int_0^\infty \hat{S}(\omega)d\omega = \sigma^2k.
\end{align}
Given the GMP driving noise variance $\sigma^2$, minimizing the total power of the bounding process is equivalent to minimizing the value of $k$. Using \eq{eq:Cond1} and \eq{eq:Cond2}, we obtain the following constrained minimization problem:
\begin{align}
    &\underset{k\in \mathbb{R}}{\min}\; k, \; \text{subject to:} \left\{\begin{array}{c}k\hat{\tau} \geq \tau_\text{max}, \\
    \frac{k}{\hat{\tau}}\geq \frac{1}{\tau_\text{min}}.\end{array}\right.
\end{align}
Figure \ref{fig:OptArea} shows that the solution is found at the intersection of the two conditions, which is found at the equality conditions. We find the following solution for non-zero correlation time-constants:
\begin{align}
    \hat{\tau} &= \sqrt{\tau_\text{min}\tau_\text{max}},  \label{eq:tauSol}\\
    k &= \frac{\tau_\text{max}}{\hat{\tau}} = \frac{\hat{\tau}}{\tau_\text{min}} = \sqrt{\frac{\tau_\text{max}}{\tau_\text{min}}}.\label{eq:kSol}
\end{align}
It can be shown that this time constant and driving noise variance multiplier are the same as the ones conjectured in \cite{GarciaCrespillo2020b_OverboundGNSSINS} using logarithmic Allan variance visualizations.

Thus, the stationary GMP model that provides the tightest bound on an actual GMP with variance $\sigma^2$ and time constant $\tau \in [\tau_\text{min}, \tau_\text{max}]$ is determined by the following time correlation and variance:
\begin{align}
    \label{eq:GMPboundDef}
	\hat{\tau} = \sqrt{\tau_\text{min}\tau_\text{max}},\hspace{3ex} 
	\hat{\sigma}^2 = \sigma^2 \sqrt{\frac{\tau_{\text{max}}}{\tau_\text{min}}}. 
\end{align}
Note that if the stationary variance of the actual GMP is only known to reside within a range $\sigma^2 \in [\sigma^2_\text{min}, \sigma^2_\text{max}]$, then it is trivial to prove that the maximum variance $\sigma^2_\text{max}$ must replace $\sigma^2$ in \eq{eq:GMPboundDef} for the GMP model to remain bounding.
\subsection{Proof for all frequencies}
The GMP model in \eq{eq:GMPboundDef} was derived considering two limiting conditions. In this subsection, we show  that this model's PSD bounds the actual PSD for all frequencies as required in \eq{eq:mainCondition}.

The GMP model in \eq{eq:GMPboundDef} has bounding PSD values at two extreme angular frequencies (for Cases 1 and 2). If the bound were not upper bounding the actual PSD at all frequencies in between, then there should be a frequency at which the bounding PSD and the actual PSD cross. By contradiction, if this were true, it would mean that:
\begin{align}
    \hat{S}(\omega) = S(\omega),\; \forall \tau \in [\tau_\text{min}, \tau_\text{max}]
\end{align}
For the GMP structure expressed in \eq{eq:GMP_PSD}, we could then write the following equation:
\begin{align}
    \frac{2\sigma^2k\hat{\beta}}{\omega^2 + \hat{\beta}^2} = \frac{2\sigma^2\beta}{\omega^2 + \beta^2}.
\end{align}
where the following change in notation is made for clarity $\beta = 1/\tau$. 
Substituting \eq{eq:tauSol} and \eq{eq:kSol} into the previous expression, rearranging, and solving for the angular frequency $\omega$, we obtain the following expression:
\begin{align}
    \omega^2 = \frac{\beta (\beta_\text{max} - \beta)}{1 - \frac{\beta}{\beta_\text{min}}}.
\end{align}
Let us test the right side of this expression for possible values of $\beta$ (and therefore $\tau$):
\begin{itemize}
    \item If $\beta = \beta_\text{max}$, the numerator is zero and therefore $\omega^2 = 0$, which relates to our first condition.
    \item If $\beta = \beta_\text{min}$, the denominator is zero and $\omega^2 \to \infty$, which relates to our second condition.
    \item If $\beta_\text{min} < \beta < \beta_\text{max}$, the denominator is negative, but the numerator is positive since $\beta$ is by definition strictly positive. This means that $\omega^2$ would be negative, which is not possible.
\end{itemize}
This means that there can be no actual GMP with time constant belonging to the range of interest that would cross the proposed bound. This proves that the GMP bound in a bound for any frequency. 

\subsection{Graphical Evaluation}
\begin{figure}[h]
    \setlength\fwidth{0.85\linewidth}
    \centering
    \input{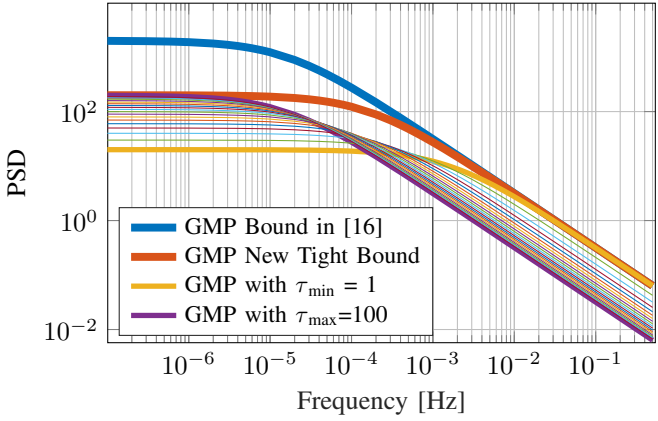}
    \caption{Power Spectral Density (PSD) of GMP with $\sigma^2 = 1$ and $\tau \in [10, 100] s$, GMP Bound in \cite{Langel2019}, and New Tight Bound.}
    \label{fig:psdNewBound}
\end{figure}

In \Fig{fig:psdNewBound}, we show the new tight stationary GMP bound, the stationary bound derived in \cite{Tupysev2009,Langel2019}, and possible realizations of the actual GMP with unit variance and time constants varying from 10 to 100 seconds.

\Fig{fig:psdNewBound} illustrates the fact that the bound in \cite{Langel2019} is looser at low frequencies than our new proposed bound, and that the new bound is the tightest possible stationary bound with a GMP structure.
%
\section{Stationary Discrete-Time GMP Model}\label{sec:stationaryBoundDis}
Sensor measurements are most often processed using digital computers. Therefore, a discrete-time LDS model is used in most practical applications:
\begin{align}
    \mb{x}_k &= \mb{\Phi}_k\mb{x}_{k-1} + \mb{w}_k, \\
    \mb{z}_k &= \mb{H}_k\mb{x}_k + \mb{v}_k,
\end{align}
where the state vector $\mb{x}$, process noise $\mb{w}$ and transition matrix $\Phi$ are discrete-time vectors and matrices.
In the same manner as in Section~\ref{sec:stationaryBoundCont}, GMP models can be incorporated by state augmentation.
A discrete-time GMP model can be written at any time epoch $n\in\mathbb{Z}\geq 0$ as:
\begin{align}
    a_{n} &= \hat{\alpha}a_{n-1} + \sqrt{\sigma^2k_d\left(1 - \hat{\alpha}^{2}\right)}w_n,
    \label{eq:GMPdisBoundModel}
\end{align}
where $\hat{\alpha} = e^{\frac{-\Delta t}{\hat{\tau}_d}}\;$ and $w_n\sim \mathcal{N}(0,1)$.

Compact expressions of $\hat{\tau}$ and $k$ for a continuous-time GMP were given in \eq{eq:GMPboundDef}. Given standard continuous to discrete transformation~\cite{Brown2012}, they can be also applied in \eq{eq:GMPdisBoundModel}, but their derivation did not consider the effect of sampling interval. In this section, we derive refined (but lengthier) expressions directly in the discrete-time domain for $\hat{\tau}_d$ and $k_d$. 

The spectral density of a discrete-time GMP can be written as~\cite{Kasdin1995_SimColorNoise}:
\begin{align}
    S_d(\omega) = \frac{\sigma^2\Delta t\left(1 - \alpha^2\right)}{1 + \alpha^2-2\alpha\cos(\omega\Delta t)}
    \label{eq:discretePSD}
\end{align}
where ${\alpha} = e^{\frac{-\Delta t}{{\tau}}}$. We want to guarantee that the spectral density of the GMP model is greater than or equal to the spectral density of the actual GMP $S_d$:
\begin{align}
    \hat{S}_d(\omega) \geq S_d(\omega),\; \forall \omega\in[0, \frac{\pi}{\Delta t}]
\end{align}
Similar to Section~\ref{sec:stationaryBoundCont}, we find the values of $\hat{\tau}_d$ and $k_d$ that produce the tightest estimation error variance bound by corresponding two limiting cases of $\omega\to 0$ and $\omega\to\frac{\pi}{\Delta t}$.  For each of these cases, we derive in Appendix~\ref{ap:disBound} the necessary constraints on $k_d$ and $\hat{\tau}_d$ and we minimize the total power of the process. \eq{eq:k_d} and \eq{eq:tau_d} in Appendix~\ref{ap:disBound} present the the resulting expressions of $k_d$ and $\hat{\tau}_d$ respectively.
\begin{figure}[h]
    \setlength\fwidth{0.8\linewidth}
    \centering
    \input{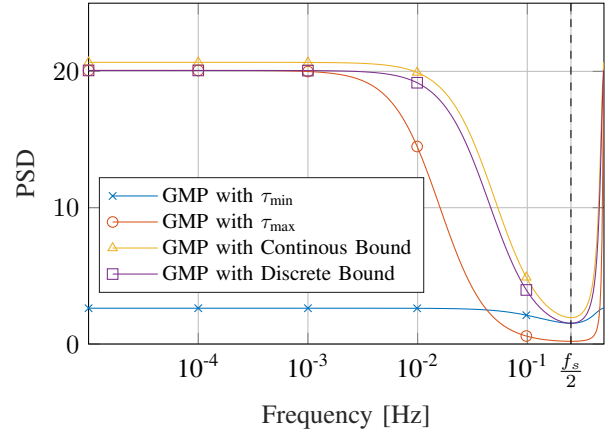}
    \caption{Discrete PSD with $\tau \in [1, 10]$ s and $f_s = \frac{1}{\Delta t} = 0.5$Hz.}
    \label{fig:discretePSD}
\end{figure}

Using these expressions, we found out that for values of $\Delta t << \tau$ the PSD of the discrete-time GMP using the continuous-time model parameters (\eq{eq:GMPboundDef}) approaches the PSD of the discrete-time model (\eq{eq:k_d}, \eq{eq:tau_d}).
Only in situations where $\Delta t$ approached or exceeded $\tau_\text{min}$, did the continuous-time GMP model become more conservative than the discrete-time model. An example situation where $\Delta t$ is greater than $\tau_\text{min}$ is shown with PSD curves in \Fig{fig:discretePSD}. It is worth noting that the GMP model with $\tau=\tau_\text{min}$ appears to be close to white (flat spectrum) due to the large sampling interval relative to $\tau$. 
Note that in most practical problems, $\Delta t$ is smaller than $\tau_\text{min}$ since otherwise, the process would be better modeled as white noise and not as a time-correlated GMP. Therefore the simpler continuous-time GMP parameter expressions are more attractive.  
%
\section{Non-Stationary Discrete-Time GMP Model}\label{sec:NonstationaryBound}
In this section, we derive an even tighter bound than in Section~\ref{sec:stationaryBoundCont} and \ref{sec:stationaryBoundDis} by considering a non-stationary process whose initial variance we must determine. This new model will match the stationary ones at steady state, but will provide a tighter estimation error variance bound during the transient period. The discrete-time GMP model can be written at any time epoch $n\in\mathbb{Z}\geq 0$ in terms of the initial condition $a_0$ as:
\begin{equation}
\begin{aligned}
    a_{n} &= \hat{\alpha}^{n}a_0 + \sqrt{\sigma^2k\left(1 - \hat{\alpha}^{2n}\right)}w_n, \\
\end{aligned}
\end{equation}
where $\hat{\alpha} = e^{\frac{-\Delta t}{\hat{\tau}}}\;$, $a_0\sim \mathcal{N}(0,\sigma_0^2)\;$, and $w_n\sim \mathcal{N}(0,1)$.
We want to find the minimum value of $\sigma_0$ that guarantees an upper bound on the estimation error variance during both the transient and steady-state periods.

The autocovariance of the non-stationary discrete-time GMP model between any two time steps $n$ and $p$ where $n\in \mathbb{Z}\geq 0, p\in \mathbb{Z}\geq 0$ and $p \geq n$ is:
\begin{align}
E[a_na_p^T] &= \hat{\alpha}^{n+p} \sigma_0^2 + \sigma^2 k (1 - \hat{\alpha}^{2n})\hat{\alpha}^{p - n}. \label{eq:DiscreteCovariance}
\end{align}
The derivation of this expression for a general non-stationary GMP can be found in Appendix~\ref{ap:NonStCovariance}.
For any $n$ and $p$ we can write the autocovariance matrix (ACM) of the GMP model as:
\begin{align}
\label{eq:Rhat}
\hat{\mathbf{R}}(n,p) = \left[\begin{array}{cc}r_{nn} & r_{np}\\r_{np}&r_{pp}\end{array}\right],
\end{align}
with
\begin{align}
r_{nn} &= \hat{\alpha}^{2n}\sigma_0^2 + \sigma^2k (1 - \hat{\alpha}^{2n}), \\
r_{np} &= \hat{\alpha}^{n+p}\sigma_0^2 + \sigma^2k (1 - \hat{\alpha}^{2n})\hat{\alpha}^{p - n}, \\ 
r_{pp} &= \hat{\alpha}^{2p}\sigma_0^2 + \sigma^2k (1 - \hat{\alpha}^{2p}).
\end{align}
The actual stationary GMP ACM is expressed as:
\begin{align}
\mathbf{R}(n,p) = \left[\begin{array}{cc}\sigma^2&\alpha^{p-n}\sigma^2\\
    \alpha^{p-n}\sigma^2&\sigma^2\end{array}\right].
\end{align}
This expression can be obtained by setting $\sigma_0^2 = \sigma^2$ and replacing $\hat{\alpha}$ with ${\alpha} = e^{\frac{-\Delta t}{{\tau}}}$ in \eq{eq:Rhat}. It is also given at the end of Appendix~\ref{ap:NonStCovariance}.

In \cite{Langel2020b}, the authors show that overbounding the estimation error is equivalent to the difference between $\hat{\mathbf{R}}(n,p)$ and $\mathbf{R}(n,p)$ being positive semidefinite.
Equivalently, we use the following notation: we want $\hat{\mathbf{R}}(n,p) \succeq \mathbf{R}(n,p) $ for any $n,p \in [0,\infty)$ and any $\tau \in [\tau_\text{min}, \tau_\text{max}]$, and with  $k$ and $\hat{\tau}$ determined in Section~\ref{sec:stationaryBoundCont} or \ref{sec:stationaryBoundDis} to ensure the tightest bound at steady-state. Therefore, we want the following inequality to be satisfied:
\begin{align}
    \left[\begin{array}{cc}r_{nn} - \sigma^2 & r_{np} - \alpha^{p-n}\sigma^2 \\
    r_{np} - \alpha^{p-n}\sigma^2& r_{pp}- \sigma^2 \end{array}\right] \succeq 0 \label{eq:transientCond1}
\end{align}
Equivalently, we want the determinant of the left-hand-side matrix in \eq{eq:transientCond1} to be greater than or equal to zero.
\begin{figure}[b] 
    \setlength\fwidth{0.8\linewidth}
    \centering
    \input{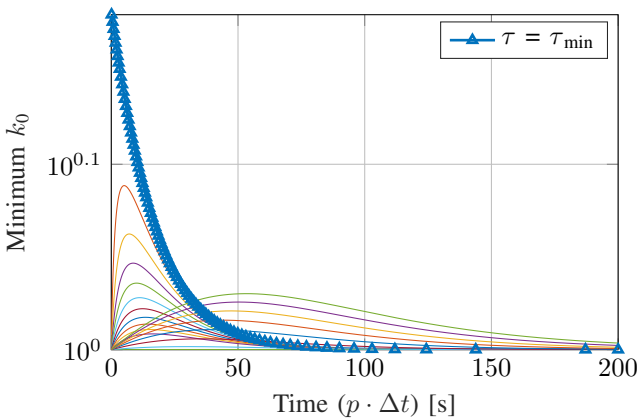}
    \caption{Minimum initial inflation factor $k_0$ with $\tau \in [10, 100]$ s and $\Delta t = 0.1$s.}
    \label{fig:conditionk0}
\end{figure}
The condition on $\sigma_0^2$ in \eq{eq:transientCond1} can be analyzed for any value of $\sigma^2, \tau, n,\ \text{and}\ p$ using computer symbolic tools (e.g., from Matlab\textsuperscript{\textregistered}). We analyzed the sensitivity of $\sigma_0^2$ to $n$ and $p$ and found out that maximum values for the condition on $\sigma_0^2$ were always obtained for $n=0$. The reason is that the effect of $\sigma_0$ on the variance of a GMP gets smaller as time passes according to \eq{eq:DiscreteCovariance}.
For $n=0$, \eq{eq:transientCond1} becomes:
\begin{align}
    \left[\begin{array}{cc}\sigma_0^2 - \sigma^2 & \hat{\alpha}^{p}\sigma_0^2 - \alpha^p\sigma^2 \\
    \hat{\alpha}^{p}\sigma_0^2 - \alpha^p\sigma^2& \hat{\alpha}^{2p}\sigma_0^2 + \sigma^2 k\left(1-\hat{\alpha}^{2p} \right)- \sigma^2 \end{array}\right] \succeq 0
    \label{eq:transientCond2}
\end{align}
which leads to the following condition on $\sigma_0^2$:
\begin{align}
    \sigma_0^2 \geq \sigma^2 \underbrace{\frac{k\left(1 - \hat{\alpha}^{2p}\right) -1 + \alpha^{2p}}{k\left(1 - \hat{\alpha}^{2p}\right) -1 - \hat{\alpha}^{2p} + 2\alpha^{p}\hat{\alpha}^p}}_{k_0} \label{eq:conditionk0}
\end{align}
Using the notation $\sigma_0^2 = \sigma^2k_0$ is convenient because we can define the inflation factor $k_0$ for the initial variance also as a function of the underlying GMP variance.

To obtain a tight bound, we want to find the minimum value of $k_0$ that is bounding for all parameter values.
\Fig{fig:conditionk0} shows the minimum value of $k_0$ for different values of $p$ (on the x-axis) and $\tau\in [10, 100]$s  (color-coded). The figure shows that the largest minimum value of $k_0$ is obtained when evaluating \eq{eq:conditionk0} for $\tau=\tau_\text{min}$ and $p=1$. 

\begin{figure}[t]
    \setlength\fwidth{0.8\linewidth}
    \centering
%
%
\definecolor{mycolor1}{rgb}{0.00000,0.44700,0.74100}%
\definecolor{mycolor2}{rgb}{0.85000,0.32500,0.09800}%
\definecolor{mycolor3}{rgb}{0.92900,0.69400,0.12500}%
\definecolor{mycolor4}{rgb}{0.49400,0.18400,0.55600}%
\definecolor{mycolor5}{rgb}{0.46600,0.67400,0.18800}%
\begin{tikzpicture}

\begin{axis}[%
width=0.951\fwidth,
height=0.7\fwidth,
at={(0\fwidth,0\fwidth)},
scale only axis,
xmode=log,
xmin=0.001,
xmax=10,
xminorticks=true,
xlabel style={font=\color{white!15!black}},
xlabel={Sampling Interval ($\Delta t$) [s]},
ymin=1,
ymax=1.9,
ylabel style={font=\color{white!15!black}},
ylabel={$\text{Minimum } k_0$},
axis background/.style={fill=white},
xmajorgrids,
xminorgrids,
ymajorgrids,
legend style={nodes={scale=0.8, transform shape}, at={(0.01,0.01)}, anchor=south west, legend cell align=left, align=left, draw=white!10!black}
]
\addplot [color=mycolor1, line width=1.0pt, mark=star, mark options={solid, mycolor1}]
  table[row sep=crcr]{%
0.001	1.51914764119409\\
0.01	1.51604222682247\\
0.1	1.48600404281608\\
1	1.26750537032218\\
10	1.00082972979033\\
};
\addlegendentry{$\tau\text{ }\in\text{ [1, 10]s}$}

\addplot [color=mycolor2, line width=1.0pt, mark=star, mark options={solid, mycolor2}]
  table[row sep=crcr]{%
0.001	1.51945924282854\\
0.01	1.51914764119409\\
0.1	1.51604222682247\\
1	1.48600404281608\\
10	1.26750537032218\\
};
\addlegendentry{$\tau\text{ }\in\text{ [10, 100]s}$}

\addplot [color=mycolor3, line width=1.0pt, mark=diamond, mark options={solid, mycolor3}]
  table[row sep=crcr]{%
0.001	1.81812729153235\\
0.01	1.81763657748093\\
0.1	1.81274857579904\\
1	1.76571335665909\\
10	1.43705043446462\\
};
\addlegendentry{$\tau\text{ }\in\text{ [10, 1000]s}$}

\addplot [color=mycolor4, line width=1.0pt, mark=triangle, mark options={solid, rotate=180, mycolor4}]
  table[row sep=crcr]{%
0.001	1.66622238522387\\
0.01	1.66223847633234\\
0.1	1.62381041349909\\
1	1.35017753564011\\
10	1.00470508873611\\
};
\addlegendentry{$\tau\text{ }\in\text{ [1, 25]s}$}

\addplot [color=mycolor5, line width=1.0pt, mark=triangle, mark options={solid, rotate=90, mycolor5}]
  table[row sep=crcr]{%
0.001	1.37857142857143\\
0.01	1.3818155661056\\
0.1	1.38196338950664\\
1	1.38193063129539\\
10	1.38161248876741\\
};
\addlegendentry{$\tau\text{ }\in\text{ [2, 10] hours}$}

\end{axis}
\end{tikzpicture}%
    \caption{Minimum initial inflation factor $k_0$ for different sample intervals and $\tau$ ranges.}
    \label{fig:samplingRateImpact}
\end{figure}
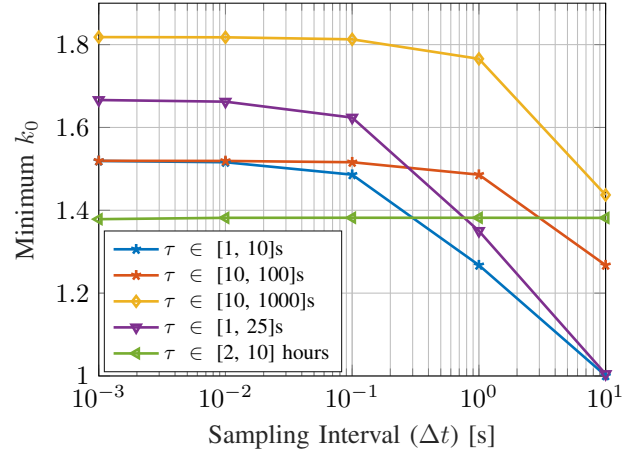

For $\tau=\tau_\text{min}$ and $p=1$, \eq{eq:conditionk0} can be rewritten as:
\begin{align}
    k_0 \geq \frac{k \left(1 - e^{-\frac{2\Delta t}{\hat{\tau}}} \right) - 1 + e^{-\frac{2\Delta t}{\tau_\text{min}}}}{k \left(1 - e^{-\frac{2\Delta t}{\hat{\tau}}} \right) - 1 - e^{-\frac{2\Delta t}{\hat{\tau}}} + 2e^{-\Delta t\left(\frac{1}{\hat{\tau}}+\frac{1}{\tau_\text{min}}\right)}}
\end{align}
It is noteworthy that the minimum value of $k_0$  is independent of $\sigma^2$, but it depends on the values and range of $\tau$ and on the sample interval $\Delta t$.
\Fig{fig:samplingRateImpact} shows the minimum value of $k_0$ versus sampling period for five examples of $\tau$-intervals.
We can see in \Fig{fig:samplingRateImpact} that if the sampling interval $\Delta t$ approaches the value of $\tau_\text{min}$, the initial variance of the non-stationary GMP model can be further relaxed through the inflation factor $k_0$. This is due to the fact that we do not need to bound the higher frequencies of the GMP since in discrete-time we are limited by the Nyquist frequency. And this margin can be used to reduce $k_0$.
\section{Example Kalman Filter Implementation}\label{sec:example}
In order to evaluate the new GMP models, we consider the motivational example described in \cite{Langel2019}, where we estimate the initial position and constant velocity of a vehicle moving along a one-dimensional trajectory. The LDS includes an augmented state to account for measurement time-correlation. The LDS is described by:
\begin{align}
    \left[\begin{array}{c}p_{0,k}\\v_k\\ \xi_k\end{array}\right] &= 
     \left[\begin{array}{ccc}1&0&0\\0&1&0\\0&0&\alpha\end{array}\right] 
     \left[\begin{array}{c}p_{0,k-1}\\v_{k-1}\\ \xi_{k-1}\end{array}\right] +
     \left[\begin{array}{c}0\\0\\ \sqrt{q_d}w_k \end{array}\right], \\
     z_k &= \left[\begin{array}{ccc}1&k\Delta t&1\end{array}\right] \left[\begin{array}{c}p_{0,k}\\v_k\\ \xi_k\end{array}\right] + \nu_k,
\end{align}
with
\begin{align}
    \begin{array}{l}\alpha = e^{\frac{-{\Delta}t}{\tau}}, \\
    q_d = \sigma_\xi^2(1 - e^\frac{-2\Delta t}{\tau}),\end{array}\hspace{1ex} \text{and}\;\begin{array}{l} w_k = \mathcal N(0,1),\\
    \nu_k = \mathcal N\left(0,\sigma_\nu^2\right), \end{array}
\end{align}
where $p_{0}$, $v$ are the initial position and speed of the vehicle and $\xi$ is the augmented state. $z$ is the position measurement. 
\begin{figure}[b]
     \setlength\fwidth{0.8\linewidth}
    \centering
    \input{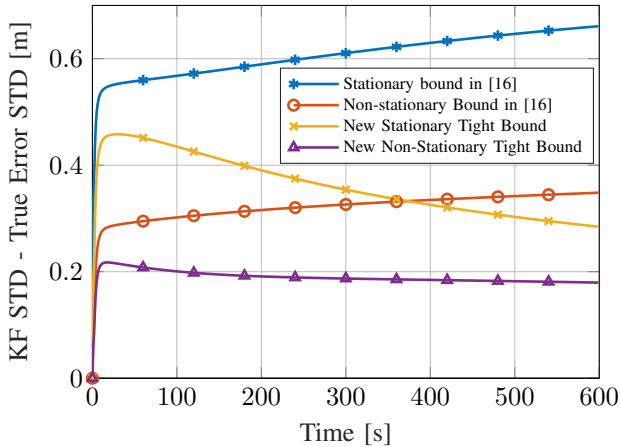}
     \caption{KF estimated error vs true error (Position) ($\tau_\text{max}=100$, $\tau_\text{min}=10$, $\tau_\text{true}=50$ and $\Delta t=1$s).}
    \label{fig:motivationExampleResults}
\end{figure}
$\tau$ is only known to be in the range $\tau \in [10,100]$s, $\sigma^2_\nu = 1$, and $\sigma^2_\xi=1$. 

In \Fig{fig:motivationExampleResults}, we show the difference between the standard deviation of the position estimated using a KF and the true estimation error standard deviation. For computation of the true estimated error of a discrete-time KF, the reader can consult the Appendix in \cite{GarciaCrespillo2020b_OverboundGNSSINS}. \Fig{fig:motivationExampleResults} displays the stationary and non-stationary GMP models derived in \cite{Langel2019} and those derived in Sections~\ref{sec:stationaryBoundCont} and \ref{sec:NonstationaryBound} of this paper. The equivalent discrete-time models of Section~\ref{sec:stationaryBoundDis} are not shown because for this example, their values are not distinguishable from the simpler stationary models of Section~\ref{sec:stationaryBoundCont}. 
Positive values of the curves mean that the GMP models all produce upper bounds on the positioning variance. 
If the simulation time was long enough, we would see positioning deviations using the non-stationary GMP models converge towards their corresponding stationary GMP models.  

We focus on the transient period.  We can see that over the first 300s of simulation time, the non-stationary GMP model in \cite{Langel2019} provides a tighter positioning deviation bound than the model in Section~\ref{sec:stationaryBoundCont}. But, as filtering approaches steady-state, our proposed stationary model in Section~\ref{sec:stationaryBoundCont} provides a tighter bound on the underlying error. 
The non-stationary GMP model in \ref{sec:NonstationaryBound} achieves the tightest positioning error bound under the paper's assumptions.

\begin{figure}[t]
     \setlength\fwidth{0.8\linewidth}
    \centering
    \input{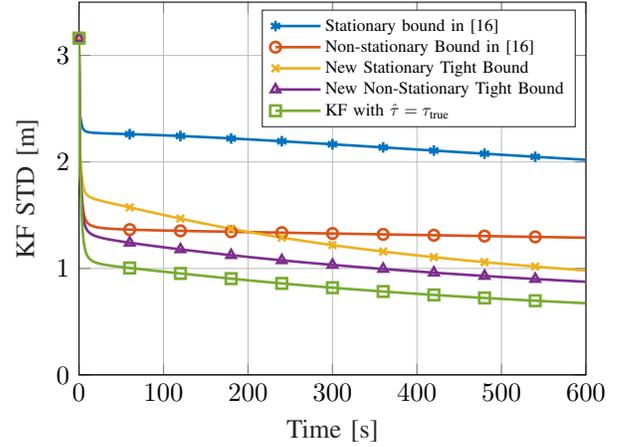}
     \caption{KF standard deviation (Position) ($\tau_\text{max}=100$, $\tau_\text{min}=10$, $\tau_\text{true}=50$ and $\Delta t=1$s).}
    \label{fig:positionAbsCov}
\end{figure}

In addition, \Fig{fig:positionAbsCov}, shows the KF positioning standard deviations for the new GMP models as compared to those in \cite{Langel2019}, and to the KF standard deviation obtained if we knew the true value of correlation time constant.  This figure illustrates the inflation in standard deviation that we endure for lack of knowledge of the actual error correlation time constant, and the tightness of the positioning variance bounds obtained using the proposed GMP models.
\section{Conclusion}\label{sec:Conclusion}
In this paper, we derived the stationary GMP model that guarantees the tightest upper bound on linear estimation error variance in the presence of uncertain time-correlated measurement and process errors with Gauss-Markov structure. The stationary model is obtained in both the continuous-time and discrete-time PSD domain. The model from both domains produced similar bound tightness in typical situations, whereas the model parameters expressions obtained in the continuous-time are less complex. The stationary model was improved upon using a non-stationary GMP which provides a tighter variance and error bound during the transient period. It was shown that these GMP models can easily be implemented in linear dynamic estimators such as Kalman filters.
%
%
%
\appendices
\section{Non-Stationary GMP Covariance over Time}\label{ap:NonStCovariance}
This Appendix provides an expression for the autocovariance of a general non-stationary discrete-time GMP between two time steps.
The first three samples of the discrete-time GMP sequence can be expressed with respect to the initial GMP sample $a_0$ as:
\begin{equation}
\begin{aligned}
    a_1 &= \alpha a_0 + \sqrt{\sigma^2 (1 - \alpha^2)}w_1, \\
    a_2 &= \alpha^2 a_0 + \alpha \sqrt{\sigma^2 (1 - \alpha^2)}w_1 + \sqrt{\sigma^2 (1 - \alpha^2)}w_2, \\
    a_3 &= \alpha^3 a_0 + \alpha^2 \sqrt{\sigma^2 (1 - \alpha^2)}w_1 + \alpha\sqrt{\sigma^2 (1 - \alpha^2)}w_2 + \\
    & +\sqrt{\sigma^2 (1 - \alpha^2)}w_3,\\
    & \vdots
\end{aligned}
\end{equation}
where
\begin{equation}
    \alpha = e^{\frac{-\Delta t}{\tau}},\;\;\;\text{and}\; w_i\sim \mathcal{N}(0,1),\,\forall i\in\mathbb{Z}>0.
\end{equation}
A general, compact form of these equations can be written for any time step $n$ as:
\begin{align}
    a_n = \alpha^n a_0 + \sqrt{\sigma^2 (1 - \alpha^2)}\sum_{i=0}^{n-1}\alpha^iw_{n-i}. \label{eq:discreteGMPwithInit}
\end{align}
Since the expected value of a GMP is zero for any time step ($E[a_n] = 0,\;\forall n\geq 0$), the autocovariance of this non-stationary process between any integer $n\in\mathbb{Z}$ and $p\in\mathbb{Z}$ time step with $p \geq n$ is:
\begin{equation}
\begin{aligned}
\label{eq:autocovPdct}
    &E[a_na_p] = \\
    &E\left[\left(\alpha^n a_0 + \sqrt{\sigma^2 (1 - \alpha^2)}\sum_{i=0}^{n-1}\alpha^iw_{n-i}\right) \right.\\
    &\left.\left(\alpha^p a_0 + \sqrt{\sigma^2 (1 - \alpha^2)}\sum_{j=0}^{p-1}\alpha^jw_{p-j}\right)\right].
\end{aligned}
\end{equation}
Using $E[a_0^2] = \sigma_0^2$ and $E[a_0w_i] = 0,\; \forall i\in\mathbb{Z}>0$, and rearranging, \eq{eq:autocovPdct} becomes:
\begin{align}
E[a_na_p] &= \alpha^n\alpha^p \sigma_0^2 + \nonumber \\
&\sigma^2 (1 - \alpha^2)\sum_{i=0}^{n-1}\sum_{j=0}^{p-1}\alpha^i\alpha^jE[w_{n-i}w_{p-j}]. \label{eq:cov2}
\end{align}
Because the driving noise $w_i$ is a white sequence, the expectation function under the double summation  in \eq{eq:cov2} is non-zero only if $n-i = p-j$, which is expressed as:
\begin{equation}
\begin{aligned}
    E[w_iw_i]  &= 1,\; \forall i>0, \\
    E[w_iw_j] &= 0,\; \text{for}\; i\neq j.
\end{aligned}
\end{equation}
Therefore we can make the change of variable:  $j = p - n + i$  to get rid of one of the two summations:
\begin{align}
E[a_na_p] &= \alpha^n\alpha^p \sigma_0^2 +
\sigma^2 (1 - \alpha^2)\sum_{i=0}^{n-1}\alpha^{2i + p -n}. \label{eq:cov3}
\end{align}
Recognizing a geometric series, \eq{eq:cov3} becomes:
\begin{align}
E[a_na_p] &= \alpha^n\alpha^p \sigma_0^2 + 
\sigma^2 (1 - \alpha^2)\frac{(\alpha^{2n} - 1)\alpha^{p - n}}{\alpha^2 - 1}, \label{eq:cov4}
\end{align}
which finally leads to:
\begin{align}
E[a_na_p] &= \alpha^{n+p}\sigma_0^2 + \sigma^2 (1 - \alpha^{2n})\alpha^{p - n},\;\; \forall p\geq n.\label{eq:cov5}
\end{align}
It is worth noting that if the process is stationary (i.e., $\sigma_0^2 = \sigma^2$), then \eq{eq:cov5} expectedly reduces to:
\begin{align}
E[a_na_p] &= \sigma^2 \alpha^{p-n},\; \forall p\geq n.\label{eq:cov6}
\end{align}
The correlation between two time steps is the same regardless of the order of indices, that is: $E[a_na_p] = E[a_pa_n]$. With this in mind, we can give an expression that does not  specify which of $n$ or $p$ is larger:
\begin{align}
E[a_na_p] &= \alpha^{n+p}\sigma_0^2 + \sigma^2 (1 - \alpha^{2\min(n,p)})\alpha^{|p - n|}.\label{eq:cov7}
\end{align}
\section{Discrete-Time GMP Parameter Expressions}\label{ap:disBound}
In this Appendix we derive a GMP model that tightly upper bounds the estimation error variance in the presence of uncertain discrete-time GMPs.
We use \eq{eq:discretePSD} defining the spectral density $\hat{S}_d$ of a discrete-time GMP model to ensure that the following inequality, where $S_d$ is the PSD of the actual GMP, is satisfied:
\begin{align}
    \hat{S}_d(\omega) \geq S_d(\omega),\; \forall \omega\in[0, \frac{\pi}{\Delta t}]
    \label{eq:disMainCond}
\end{align}
We analyze two limiting situations:
\begin{itemize}
    \item Case 1: We consider the case where $\omega \to 0$. In this case, the cosine in denominator of \eq{eq:discretePSD} is equal to one and \eq{eq:disMainCond} can be written as:
    \begin{align}
\frac{\sigma^2k_d\Delta t\left(1 - \hat{\alpha}^2\right)}{1 + \hat{\alpha}^2-2\hat{\alpha}} \geq \frac{\sigma^2\Delta t\left(1 - \alpha^2\right)}{1 + \alpha^2-2\alpha}
\label{eq:disCond1}
    \end{align}
Recognizing identities, dividing the numerator and denominator by $\left(1 - \hat{\alpha}\right)$ on the left-hand side, and by $\left(1 - {\alpha}\right)$ on the right-hand side, and dividing both sides by $(\sigma^2 \Delta t)$, \eq{eq:disCond1} becomes:
\begin{align}
    \frac{k_d(\hat{\alpha} + 1)}{(1 -\hat{\alpha})} \geq \frac{(\alpha + 1)}{(1 - \alpha)}
    \label{eq:disCond2}
\end{align}
    This condition must hold for any value of $\alpha$ which is a function $\tau \in [\tau_\text{min},\tau_\text{max}]$. The maximum value of the right-hand side of \eq{eq:disCond2} is achieved when $\alpha = \alpha_\text{max}$ so that $\tau=\tau_\text{max}$ must be chosen for this condition.
    
Solving for $k_d$, we end up with:
\begin{align}
    \text{Constraint 1:}\;\;\; k_d \geq \frac{(\alpha_\text{max} + 1)(1 - \hat{\alpha})}{(1 - \alpha_\text{max})(\hat{\alpha} + 1)}
\end{align}
where we used the fact that $0 < \hat{\alpha} <1$.
    
    \item Case 2: We consider the situation where $w \to \frac{\pi}{\Delta t}$. In this case, the cosine in \eq{eq:discretePSD} is equal to $-1$ and \eq{eq:disMainCond} becomes:
        \begin{align}
\frac{\sigma^2k\Delta t\left(1 - \hat{\alpha}^2\right)}{1 + \hat{\alpha}^2 + 2\hat{\alpha}} \geq \frac{\sigma^2\Delta t\left(1 - \alpha^2\right)}{1 + \alpha^2 + 2\alpha}
    \end{align}
Proceeding in the same way as in Case 1, we find that the most limiting situation is when $\alpha=\alpha_\text{min}$ and therefore, the second condition can be written as:
\begin{align}
    \text{Constraint 2:}\;\;\; k_d \geq \frac{(1 - \alpha_\text{min})(\hat{\alpha} + 1)}{(\alpha_\text{min} + 1)(1 - \hat{\alpha})}
\end{align}
\end{itemize}
The two conditions form a trade space similar to the one shown in \Fig{fig:OptArea}. We want to minimize the value of $k_d$, which directly impacts the total variance of the process, under the constraints of the above two conditions. 
The solution for $k_d$ is found to be:
\begin{align}
    k_d = \frac{\sqrt{e^{\frac{-2\Delta t}{\tau_\text{min}}}e^{\frac{-2\Delta t}{\tau_\text{max}}}-e^{\frac{-2\Delta t}{\tau_\text{max}}} - e^{\frac{-2\Delta t}{\tau_\text{min}}} + 1}}{e^{\frac{-\Delta t}{\tau_\text{max}}} - e^{\frac{-\Delta t}{\tau_\text{min}}} + e^{\frac{-\Delta t}{\tau_\text{min}}}e^{\frac{-\Delta t}{\tau_\text{max}}} -1}
    \label{eq:k_d}
\end{align}
and the solution for $\hat{\tau}_d$ is the following:
\begin{equation}\label{eq:tau_d}
\begin{aligned}
    \hat{\tau}_d = &\Delta t \left\{\ln\left(\sqrt{\left(e^{\frac{-\Delta t}{\tau_\text{min}}} -1\right)\left(e^{\frac{-\Delta t}{\tau_\text{min}}}+1\right)}\cdot \right.\right. \\
    &\left.\sqrt{\left(e^{\frac{-\Delta t}{\tau_\text{max}}} -1\right)
    \left(e^{\frac{-\Delta t}{\tau_\text{max}}}+1\right)} + e^{\frac{-\Delta t}{\tau_\text{min}}}e^{\frac{-\Delta t}{\tau_\text{max}}} + 1\right) \\
    &\left.-\ln\left({e^{\frac{-\Delta t}{\tau_\text{min}}} + e^{\frac{-\Delta t}{\tau_\text{max}}}}\right)\right\}^{-1}
\end{aligned}
\end{equation}

Because we assume that the time correlation's structure is that of a GMP, it can be shown, similar to the continuous case, that this solution not only holds at the limit cases, but also for any angular frequency $\omega \in [0,\frac{\pi}{\Delta t}]$.
\ifCLASSOPTIONcaptionsoff
  \newpage
\fi
%
%
%
%
%
%
\bibliographystyle{IEEEtran}
\bibliography{References/Overbounding.bib, References/references_PLANS2020.bib, References/robust.bib}
%
%
%
%
%
\begin{IEEEbiography}
[{\includegraphics[width=1in,height=1.25in,clip,keepaspectratio]{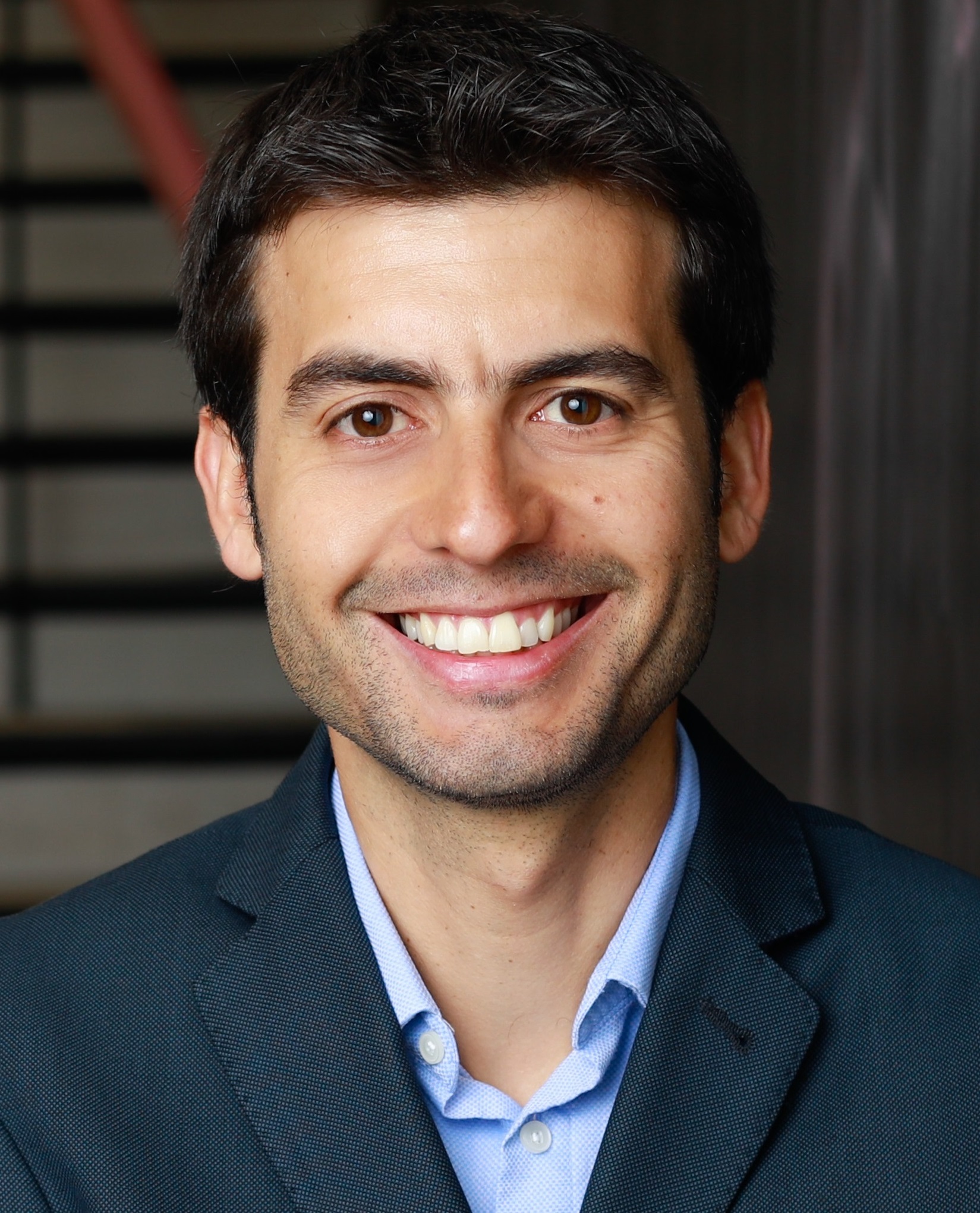}}]{Omar Garc\'ia Crespillo}
holds a M.Sc. in Telecommunication Engineering from the University of Malaga in Spain. In 2013, he joined the Navigation department of the German Aerospace Center (DLR) where his current field of research includes multi-sensor fusion algorithms, GNSS, inertial sensors and integrity monitoring for safe ground and air transportation systems. Since 2015, he is also a PhD student at the Swiss Federal Institute of Technology (EPFL) in Lausanne.
\end{IEEEbiography}
\begin{IEEEbiography}[{\includegraphics[width=1in,height=1.25in,clip,keepaspectratio]{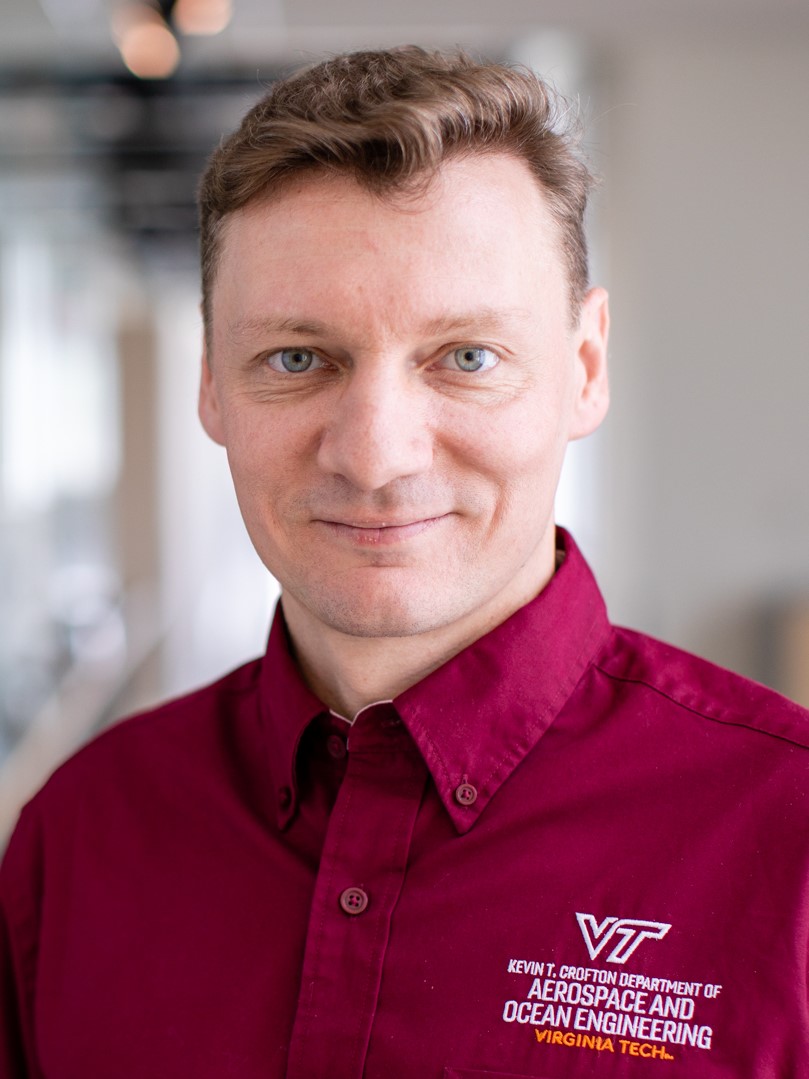}}]{Mathieu Joerger}
(M'14–SM'18) received the `Dipl\^ome d'Ingenieur' (Master degree)
in Mechatronics from the National Institute of Applied Sciences, Strasbourg, France, the
M.S. and Ph.D. degree in mechanical and aerospace engineering from the Illinois Institute
of Technology (IIT), Chicago, IL, USA.
He is currently an Assistant Professor of Aerospace and Ocean Engineering with
Virginia Tech, Blacksburg, VA, USA, working on multisensor integration, on
sequential fault-detection for multiconstellation navigation systems, and on receiver autonomous integrity monitoring for automotive and aviation applications.
Prior to joining VT, he was an Assistant Professor at the University of Arizona, Tucson, AZ, and a Reseacrh Assistant Professor at IIT.
Dr. Joerger is currently the Technical Editor of Navigation for IEEE Transactions on Aerospace \& Electronic Systems.  He was the recipient of the Institute of Navigation (ION) Bradford Parkinson Award
in 2009, and ION Early Achievement Award in 2014.
\end{IEEEbiography}
\begin{IEEEbiography}[{\includegraphics[width=1in,height=1.25in,clip,keepaspectratio]{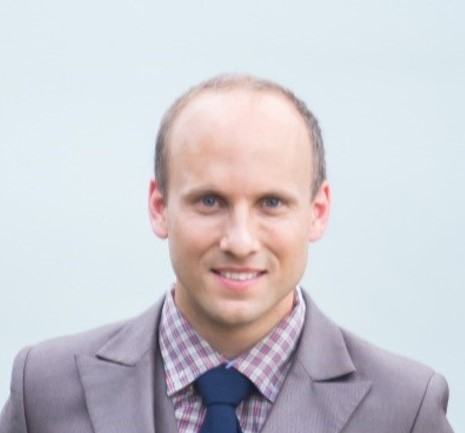}}]{Steve Langel} received a Ph.D. in mechanical and aerospace engineering from the Illinois Institute of Technology (IIT), Chicago, IL, USA. He is currently a lead signal processing engineer at The MITRE Corporation, Bedford, MA, USA, focusing on the development of multi-sensor navigation and fault detection algorithms. He is also pursuing research in robust estimation algorithms for safety-critical applications. Steve has been a member of the MITRE technical staff since 2014.
\end{IEEEbiography}
%
%
%
%
%
\end{document}